\documentclass{appolb}
\usepackage{epsfig}

\begin{document}
\pagestyle{plain}
\title{Is Quark Saturation Related to the Pauli Principle?\thanks{Dedicated to Professor Jan Kwiecinski in celebration of his 65th birthday}}
\author{A.H. Mueller\footnote{This research is supported in part by the US Department of Energy.}
\address{Department of Physics, Columbia University\\
New York, New York 10027}}

\maketitle
\begin{abstract}Below the saturation momentum the sea quark occupation number reaches a pure number, independent of any parameters in QCD, reminiscent of a Pauli Principle result.  We argue, however, that the Pauli Principle plays no role in this result.
\end{abstract}
  
\section{Introduction}
The idea of partonic saturation[1] is that the occupation numbers of sea quarks and gluons in the light-cone wavefunction of a fast hadron or of a large nucleus do not grow so rapidly as the total number of partons but reach a limiting distribution, saturation.  In the case of quarks the statement is that below some particular momentum, the saturation momentum $Q_s,$ quark occupations reach a pure, geometric, number independent of any dynamics.  Above the saturation momentum quark occupation numbers in the light-cone wavefunction are small and calculable using perturbative techniques.  The saturation result is given below in (11), (12) and (13).

Eq.(13) looks very much like the result of a Pauli Principle constraint where all levels well below $Q_s$ are fractionally filled while all levels well above $Q_s$ have small occupation numbers.  The fact that levels lying below $Q_s$ cannot have large occupation numbers is guaranteed by the Pauli Principle.

The purpose of the present paper is to examine whether or not it is the Pauli Principle which is forcing quark occupation numbers not to be large.  Our conclusion is that the Pauli Principle does not play an important role in the emergence of (13).

In Sec.2 a review of quark saturation in a high-energy light-cone wavefunction is given. The McLerran-Venugopalan model[2-4] is used for purposes of illustration because the calculation can be carried to completion with little technical difficulty.  Because of the use of a technical result, (7), which may not have complete generality and because the uncertainty principle is used in going from (11) to (12) the result (13) should be taken as a ``rough'' result.  Occupation numbers, for momenta below $Q_s,$ reach a limit, but that limit  may not be exactly the ${1\over \pi}$ given in (13) which followed from the uncertainty principle and the McLerran-Venugopalan model.

In Sec.3, we show how to deal with the Pauli Principle when doing light-cone perturbation theory calculations.  The Pauli Principle shows up as a cancellation of two separate terms when evaluating the overlap of wavefunctions where two or more quark-antiquark pairs are present.

In Sec.4, we argue that multi-quark-antiquark loop contributions cancel among themselves in the small  \ x\ part of a light-cone   wavefunction.  This leaves only the one-loop contribution which has no constraint from the Pauli Principle.  It should be emphasized that the one-loop contribution, that is a contribution linear in $N_f,$ can show up in the light-cone wavefunction as many quark-antiquark pairs.  Thus all the graphs in Fig.6 are one-loop, linear in $N_f,$ but the different contributions, when viewed as light-cone wavefunction contributions, have differing number of pairs.

\section{A brief review of quark saturation}

It is very simple to derive the form of a saturated quark distribution for a large nucleus. The form obtained[5-7] is expected to be valid also for protons and other hadrons at small x, but the argument is especially simple for a large nucleus where the sea quark distribution can be viewed as the sea distribution in the background field (the Weizs\"acker-Williams field) of the large nucleus.  However, calculationally it is simplest to determine the sea disribution by scattering a high-energy virtual photon on a large nucleus at rest.  In this picture the virtual photon splits into a quark-antiquark pair before reaching the nucleus. The pair then scatters on the nucleons of the nucleus through one and two-gluon exchanges and emerges as a  pair of produced quark jets at the back end of the nucleus.  The momentum distribution of, say, the quark jet is identical to the momentum distribution of a sea quark in the infinite momentum wavefunction of the nucleus,  at least in a particular light-cone gauge, the Kovchegov gauge[4], which eliminates final state interactions.

\begin{center}
\begin{figure}
\epsfbox[0 0 218 71]{FIG.1.1084.eps}
\\
\centerline{FIG.1}
\end{figure}
\end{center}

The scattering of a high-energy virtual photon on a nucleus is schematically pictured in Fig.1 where some sample interactions of the dipole with nucleons in the nucleus is illustrated.  The formula for the sea quark plus antiquark distribution is[8]

$$
e_f^2{d(xq_f+ x \bar{q}_f)\over d^2bd^2\ell} = {Q^2\over 4\pi^2\alpha_{em}}
\int {d^2x_1d^2x_2\over 4\pi^2}\int_0^1dz{1\over 2}
$$

\begin{equation}
\cdot \sum_\lambda \psi_{T\lambda}^{f\ast}(\underline{x}_2,z,Q)\psi_{T\lambda}^f(\underline{x}_1,z,Q) e^{-i\underline{\ell}\cdot(\underline{x}_1-\underline{x}_2)}[S^\dag (\underline{x}_2)-1][S(\underline{x}_1)-1].
\end{equation}

\noindent where $\underline{x}_2$ is the transverse coordinate of the measured quark (or antiquark) in the complex conjugate amplitude and $\underline{x}_1$ is the analogous quantity in the amplitude.   $S(\underline{x})$ is the $S-$matrix for the scattering of a dipole of size $\underline{x}$ in the amplitude while $S^\dag$ represents the scattering in the complex conjugate amplitude.  The transverse wavefunction of the virtual photon to go into a quark-antiquark pair is

$$
\psi_{T\lambda}^f(\underline{x}, z, Q) = \{{\alpha_{em}N_c\over 2\pi^2} z(1-z)[z^2+ (1-z)^2]Q^2\}^{1/2}
$$

\begin{equation}
\cdot e_f K_1({\sqrt{Q^2\underline{x}^2z(1-z)}}\ ){\underline{\epsilon}^\lambda\cdot\underline{x}\over \vert\underline{x}\vert}
\end{equation}

\noindent where $\underline{\epsilon}^\lambda$ is the polarization vector of the photon whose direction of motion has been chosen to be along the $z-$axis.

Now $\vert S(\underline{x})\vert^2$ is the probability that the dipole not have an inelastic interaction as it passes through the nucleus.  We can write

\begin{equation}
\vert S(\underline{x})\vert^2 = e^{-L/\lambda}
\end{equation}

\noindent where $L=2{\sqrt{R^2-b^2}}$ is the length of nuclear material that the dipole tranverses at impact parameter \ b\ for a spherical nucleus of radius  \ R,\ and $\lambda$ is the mean free path for inelastic dipole-nucleon interactions.  Using

\begin{equation}
\lambda =[\rho\sigma]^{-1}
\end{equation}

\noindent with $\rho$ the nuclear density one has

\begin{equation}
S(\underline{x}, b) = exp[-2{\sqrt{R^2-b^2}}\ \rho\sigma(\underline{x})/2]
\end{equation}

\noindent if we suppose \ $S$\ is purely real.  Detailed calculation gives

\begin{equation}
\sigma(\underline{x}) = {\pi^2\alpha\over N_c} x G(x, 1/\underline{x}^2)\underline{x}^2.
\end{equation}

We need one technical result in order to simplify (1) into a usable form. That result is

\begin{equation}
S^\dag(\underline{x}_2) S(\underline{x}_1) = S(\underline{x}_1-\underline{x}_2)
\end{equation}

\noindent which says that when arbitrary interactions are allowed in the initial and final state the interactions with the spectator antiquark (or quark), at $\underline{x} = 0$ in our notation, cancel completely with the result that $S^\dag S$ looks like the interaction of a dipole of  a size $\underline{x}_1-\underline{x}_2$ with the nucleons of the nucleus.  Thus we may write

$$
e_f^2{d(xq_f+ x \bar{q}_f)\over d^2bd^2\ell} = {Q^2\over 4\pi^2\alpha_{em}} \int {d^2x_1d^2x_2\over 4\pi^2}\int_0^1 dz{1\over 2}\sum_\lambda \psi_{T\lambda}^{f\ast}
$$

\begin{equation}
\cdot \psi_{T\lambda}^f e^{-i\underline{\ell}\cdot (\underline{x}_1-\underline{x}_2)}\left[1+e^{-(\underline{x}_1-\underline{x}_2)^2\bar{Q}_s^2/4}-e^{-\underline{x}_1^2\bar{Q}_s^2/4}-e^{-\underline{x}_2^2\bar{Q}_s^2/4}\right]
\end{equation}

\noindent with

\begin{equation}
\bar{Q}_s^2 = {C_F\over N_c}Q_s^2
\end{equation}

\noindent and

\begin{equation}
Q_s^2 = {8\pi^2\alpha N_c\over N_c^2-1} \rho{\sqrt{R^2-b^2}}\  x G.
\end{equation}

Now using (2) it is straightforward to evaluate (8) when $\underline{\ell}^2/\underline{Q}_s^2 \ll 1$ with the result[8]

\begin{equation}
{d(xq_f+x\bar{q}_f)\over d^2bd^2\ell} = {N_c\over 2\pi^4}
\end{equation}

\noindent We may, roughly, turn this into an ocupation number by using $dy = {d\ell_z\over \ell_z} \simeq d\ell_z db_z$ so that the occupation number for sea quarks \underline{or} antiquarks becomes

\begin{equation}
f_q \simeq {(2\pi)^3\over 2\cdot 2\cdot N_c} {d(xq_q+ x\bar{q}_q)\over d^2\ell d^2b} \simeq {(2\pi)^3\over 2\cdot 2\cdot N_c}
 {dN_{q+\bar{q}}\over d^3\ell d^3b}
\end{equation}

\noindent where the division by the two 2's comes from the counting of spin and particle and antiparticles states.  Thus, when $\ell_\perp/\bar{Q}_s\ll 1,$

\begin{equation}
f_q \simeq {1\over \pi}.
\end{equation}

\noindent Because we have used the uncertainty principle to go to a three-dimensional distribution, (13) should be considered as a rough result.  Nevertheless, the result (13) looks surprisingly like a Pauli Principle result, the occupation number of a particular species of fermion has a maximum value less than 1.  In the following section we shall review the Pauli Principle in the context of a light-cone wavefunction and try to determine whether (13) is related to such a principle.

\section{The Pauli Principle}

In light-cone quantization one introduces annihilation operators for quarks, $b_r(\underline{p}, p_+),$ and for antiquarks, $d_r(\underline{p}, p_+),$ along with the corresponding creation    operators.  The free field anticommutation relations are

\begin{equation}
\{b_{r^\prime}(\underline{p}^\prime, p_+^\prime), b_r^\dag(\underline{p}, p_+)\} =\delta_{rr^\prime}\delta(\underline{p}-\underline{p}^\prime)\delta(p_+-p_+^\prime)
\end{equation}

\begin{equation}
\{d_{r^\prime}(\underline{p}^\prime, p_+^\prime), d_r^\dag(\underline{p}, p_+)\} = \delta_{rr^\prime}\delta(\underline{p}-\underline{p}^\prime) \delta(p_+-p_+^\prime)
\end{equation}

\noindent where $r,$ and $r^\prime$ label spin while $\underline{p}$ and $p_+$ are transverse and light-cone components of the momentum.  The anticommutation relations (14) and (15) guarantees that any state vector in the Fock space obeys the Pauli Exclusion Principle if such a state is constructed \underline{explicitly} using the creation operators as $b^\dag$ and $d^\dag.$

However, when using light-cone perturbation theory it is not convenient to enforce the Pauli Principle in this explicit way.  For example, in the McLerran-Venugopalan model the valence quarks in the nucleus are viewed as sources for non-Abelian Weiz\"acker-Williams fields which then may create (virtual) quark-antiquark pairs as illustrated for the case of two pairs in Fig.2, where the direction of the arrows indicates whether the fermion is a quark or an antiquark.

\begin{center}
\begin{figure}
\epsfbox[0 0 257 144]{FIG.2.1084.eps}\\
\\
\centerline{FIG.2}
\end{figure}
\end{center}

\noindent The state of two quark-antiquark pairs shown in Fig.2 is not manifestly the zero state when, say,  $(\underline{p}_1, p_{1+})=(\underline{p}_2, p_{2+}).$  However, the zero appears when we consider the overlap (norm) of the state with itself in which case there are two distinct terms when $p_1=p_2$ as shown in Fig.3 where, for simplicity of illustration,  we omit the gluonic connections.  We have, as well, suppressed the spin labels. In Fig.3 we have taken $p_1=p_2=p.$  If $p_1\not= p_2$ then the graph shown in Fig.3b is absent.  Now the terms in Figs.3a and 3b  are identical.  Since the momenta, $p, p_3, p_4$ in the wavefunction and complex conjugate wavefunction are exactly identical in the amplitude and complex conjugate amplitude for the two graphs there is no difference in their values.  We impose a difference by assigning an extra $(-1)$ to the graph shown in Fig.3b to make the light-cone perturbation theory calculation agree with the Feynman graph calculation. In addition the graph of Fig.3a has generically a counting factor $(N_f)^2$ while that of Fig.3b gets only a $N_f$ factor.  But these factors are exactly right to give a cancellation between the two graphs in case the two $p-$lines in Fig.3a have identical colors, spins and flavors.  If the two $p-$lines do not have identical colors, spins and flavors there will not be a cancellation.  This is how the Pauli Principle emerges in doing light-cone gauge perturbation theory calculations.  It is straightforward to see that the identical cancellation continues to occur also for scattering processes and when more than two quark-antiquark pairs are present in the light-cone wavefunction.

\begin{center}
\begin{figure}
\epsfbox[0 0 263 93]{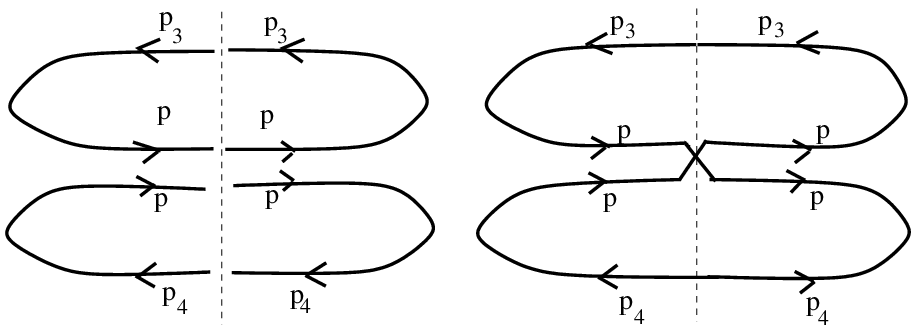}\\
\flushleft{FIG.3a}\centerline{ Fig.3b}
\end{figure}
\end{center}

\section{Cancellation of multi-fermionic loops}

In this section we are going to show that, when used to determine the quark number density, the graphs of Fig.3a cancel with similar two fermion loop graphs, but where one of the fermion loops occurs at times before the quark density is measured.  We shall demonstrate this cancellation only in a simple example and then comment on the circumstances where one can expect the cancellation to be valid.  To that end consider the graphs in Fig.4.  The two solid lines at the top of the graphs are two valence quarks, coming from different nucleons, which serve as sources in the McLerran-Venugopalan model.  The sea quark $p,$ marked with an $x$ in the graph is the quark given by, say, (12) or (13).  However, the sum of the contributions  of the various graphs of Fig.4 gives a zero result for the ``observed'' quark density because of the cancellation between the term where the second quark loop appears in the state at the time of observation of the quark  \ $p$\   and the other terms where the second quark loop appears only at earlier times in the amplitude or complex conjugate amplitude.  This cancellation is the same as the well-known cancellation shown in Fig.5 representing the fact that the probability that the system has  a quark-antiquark component in its wavefunction due to virtual pair emission must be compensated by a decrease in the probability of the other components of the wavefunction.

\begin{center}
\begin{figure}
\epsfbox[0 0 376 94]{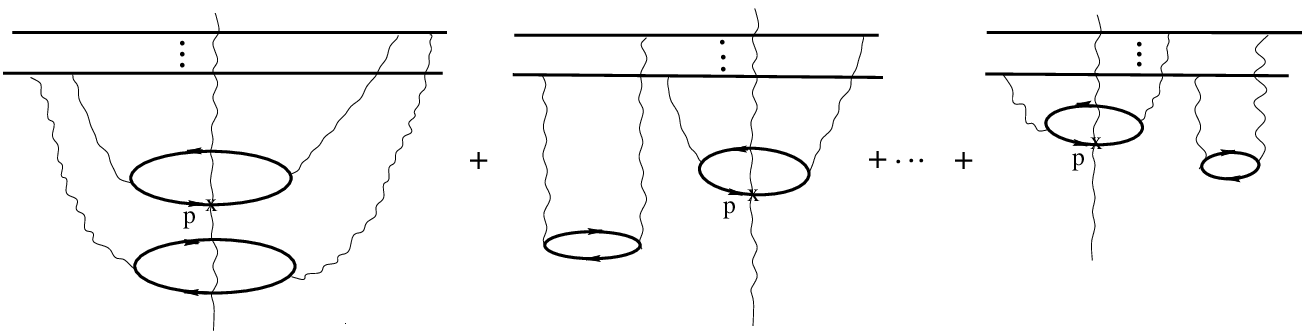}
\\
\centerline{FIG.4}
\end{figure}
\end{center}

\noindent Such a cancellation is certainly not a completely general phenomenon.  For example, gluonic interactions between the two fermionic loops of Fig.4 would destroy the cancellation by correlating the fermionic loops.  However, such interactions are higher order in $\alpha$ with no enhancing logarithmic factors, and so can be expected to be small.  If the two fermionic loops come from separate BFKL evolutions or from sources in the McLerran-Vengopalan model the cancellation again will be effective. So long as the fermion pairs are at small values of \ $x$\ we believe they will not be strongly correlated and hence that the quark density in the light-cone wavefunction is determined from single-loop graphs of the form shown in Fig.6.  For these graphs the Pauli Principle does not give constraints on the value of the quark density.  Thus we believe that it is not the Pauli Principle which leads to limits on the sea quark density in the light-cone wavefunction of a fast hadron or of  a large nucleus.

\begin{center}
\begin{figure}
\epsfbox[0 0 342 50]{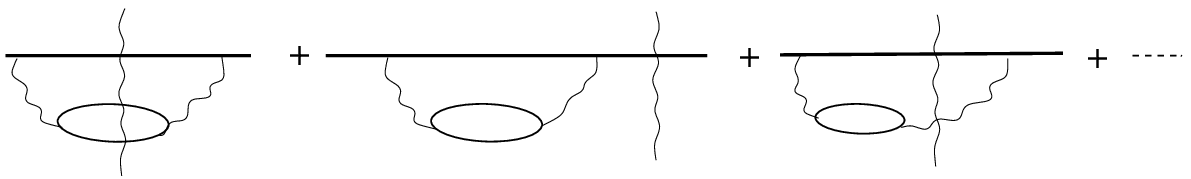}
\\
\centerline{FIG.5}
\end{figure}
\end{center}

\begin{center}
\begin{figure}
\epsfbox[0 0 353 92]{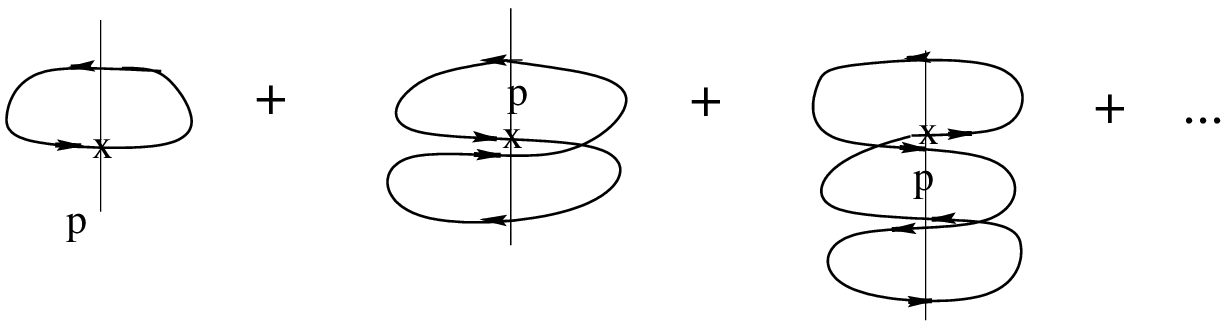}\\
\\
\centerline{FIG.6}
\end{figure}
\end{center}

\end{document}